# Radiation pressure and the linear momentum of light in dispersive dielectric media

**Masud Mansuripur**

*Optical Sciences Center, The University of Arizona, Tucson, Arizona 85721*
*masud@optics.arizona.edu*

**Abstract**: We derive an exact expression for the radiation pressure of a quasi- monochromatic plane wave incident from the free space onto the flat surface of a semi-infinite dielectric medium. In order to account for the total optical momentum (incident plus reflected) that is transferred to the dielectric, the mechanical momentum acquired by the medium must be added to the rate of flow of the electromagnetic momentum (the so-called Abraham momentum) inside the dielectric. We confirm that the electromagnetic momentum travels with the group velocity of light inside the medium. The photon drag effect in which the photons captured in a semiconductor appear to have the Minkowski momentum is explained by analyzing a model system consisting of a thin absorptive layer embedded in a transparent dielectric.

**OCIS codes**: (260.2110) Electromagnetic theory; (140.7010) Trapping.

**References**

1. M. Mansuripur, "Radiation pressure and the linear momentum of the electromagnetic field," Opt. Express **12**, 5375-5401 (2004), http://www.opticsexpress.org/abstract.cfm?URI=OPEX-12-22-5375.
2. J. P. Gordon, "Radiation forces and momenta in dielectric media," Phys. Rev. A **8**, 14-21 (1973).
3. J. D. Jackson, *Classical Electrodynamics*, 2$^{nd}$ edition (Wiley, New York, 1975).
4. R. Loudon, "Theory of the radiation pressure on dielectric surfaces," J. Mod. Opt. **49**, 821-838 (2002).
5. Y. N. Obukhov and F. W. Hehl, "Electromagnetic energy-momentum and forces in matter," Phys. Lett. A **311**, 277-284 (2003).
6. M. Mansuripur, A. R. Zakharian, and J. V. Moloney, "Radiation pressure on a dielectric wedge," accepted for publication, Opt. Express, 2005.
7. A. F. Gibson, M. F. Kimmitt, and A. C. Walker, "Photon drag in Germanium," Appl. Phys. Lett. **17**, 75-77 (1970).
8. R. Loudon, S. M. Barnett, and C. Baxter, "Theory of radiation pressure and momentum transfer in dielectrics: the photon drag effect," to appear in 2005.

## 1. Introduction

In a previous paper [1] we showed that the momentum density *p* of a plane electro-magnetic wave inside a dispersionless dielectric medium may be expressed as the average of the Minkowski and Abraham momentum densities [2], namely,

$$\boldsymbol{p} = \tfrac{1}{4}\, Real\, (\boldsymbol{E} \times \boldsymbol{H}^*)/c^2 + \tfrac{1}{4}\, Real\, (\boldsymbol{D} \times \boldsymbol{B}^*). \tag{1a}$$

Here the complex amplitudes of the electric and magnetic fields within the medium of refractive index *n* are denoted by **E** and **H**, respectively; $\boldsymbol{B} = \mu_o \boldsymbol{H}$, and $\boldsymbol{D} = \varepsilon_o \boldsymbol{E} + \boldsymbol{P} = \varepsilon_o \varepsilon \boldsymbol{E}$, where **P** is the polarization density induced in the medium by the local *E*-field. $\varepsilon_o$ is the permittivity and $\mu_o$ the permeability of free space; $\varepsilon = n^2$ is the relative permittivity of the dielectric material [3]. We derived Eq. (1a) by a direct application of the Lorentz law of force to bound charges and bound currents within the medium – a method that has been the subject of other recent studies as well [4,5]. It was concluded that the light carries its own electro-magnetic momentum inside the dielectric, while an additional momentum is transferred to the medium in the form of mechanical force. Equation (1a) may be rewritten as follows:

$$\boldsymbol{p} = \tfrac{1}{2} Real\,(\boldsymbol{E} \times \boldsymbol{H}^*)/c^2 + \tfrac{1}{4} Real\,(\boldsymbol{P} \times \boldsymbol{B}^*). \qquad (1b)$$

In the above equation, the first term is the Abraham momentum density of the field, while the second term is the mechanical momentum density imparted to the medium. (If the coefficient of the second term were ½ instead of ¼, the total momentum density $\boldsymbol{p}$ would have been equal to the Minkowski momentum.) The electromagnetic and mechanical momenta of the light inside the dielectric are *not* decoupled from each other. This fact is better appreciated if one observes, for instance, that the same beam of light, upon emerging into the free-space at the exit facet of a dielectric slab, recovers its total initial momentum (i.e., the momentum it possessed before entering the slab) by re-converting the mechanical momentum (manifested in the motion of the medium) to electromagnetic momentum [1]. If, for simplicity's sake, we assume that the entrance and exit facets of the dielectric slab are anti-reflection coated, then, upon transmission, the emerging beam's momentum will be identical to the momentum it possessed before entering the slab; in other words, the (partial) conversion of the beam's momentum into mechanical form that takes place while the beam passes through the slab, is fully reversed when the beam leaves the slab and returns to the free space. Another example of the "connectedness" of the electromagnetic and mechanical momenta was provided in [6], where the radiation pressure on a dielectric wedge and its surrounding liquid was found to arise from the *total* momentum of the beam as opposed to, say, from one or the other of its constituents.

The present paper extends the results of our previous work to the case of light beams that propagate in dispersive dielectrics. We show that the earlier results obtained for non-dispersive media remain valid if the Abraham momentum is assumed to travel with the group velocity $V_g = c/(n + n'f)$ inside the dielectric. ($f$ is the optical frequency; $n' = dn/df$ is the derivative of the refractive index $n$.) Our new expression for the mechanical momentum density reverts to the old expression, $\tfrac{1}{4} Real\,(\boldsymbol{P} \times \boldsymbol{B}^*)$, in the limit of $n' \to 0$, i.e., when the medium becomes dispersionless. Finally, to resolve the discrepancy between the theory and certain experiments in which the light appears to possess the Minkowski momentum, we propose a model system for analyzing the photon drag effect observed in certain bulk semiconductors.

## 2. Superposition of two plane waves in free space

Figure 1 shows a beam of light consisting of two equal-amplitude plane-waves of differing frequencies $f_1$ and $f_2$, incident on a semi-infinite dielectric medium of refractive index $n(f)$. The beam is linearly polarized, having its *E*-field along the *x*-axis and *H*-field along the *y*-axis. The field amplitudes in free space are given by

$$E_x(z, t) = E_o \sin\{2\pi f_1 [(z/c) - t]\} - E_o \sin\{2\pi f_2 [(z/c) - t]\} \qquad (2a)$$

$$H_y(z, t) = (E_o/Z_o) \sin\{2\pi f_1 [(z/c) - t]\} - (E_o/Z_o) \sin\{2\pi f_2 [(z/c) - t]\} \qquad (2b)$$

Here $Z_o = \sqrt{\mu_o/\varepsilon_o}$ is the free-space impedance, and $c = 1/\sqrt{\mu_o \varepsilon_o}$ is the speed of light in vacuum. The traveling wave is readily seen to be a sinusoid of frequency $f = \tfrac{1}{2}(f_1 + f_2)$, modulated with another (envelope) sinusoid of frequency $\Delta f = f_2 - f_1$, exhibiting a beat period $T = 1/\Delta f$, and traveling (in free-space) with the speed $c$.

The Poynting vector $\boldsymbol{S} = \boldsymbol{E} \times \boldsymbol{H}$ has a component only along the *z*-axis, $S_z(z, t) = E_x(z, t) H_y(z, t)$. For a fixed value of *z*, if $S_z(z, t)$ is averaged over the time interval $T$, and if terms of order $\Delta f$ and higher are neglected, the time-averaged Poynting vector will become independent of the coordinate *z*, and will be given by

$$\langle S_z(z, t)\rangle = E_o^2/Z_o \tag{3}$$

(The above equation will be exact, i.e., terms in $\Delta f$ and higher order will be absent, if $f = N\Delta f$, i.e., if the center frequency happens to be an integer-multiple of $\Delta f$.) Each participating frequency thus contributes its own (time-averaged) Poynting vector $\langle S_z\rangle = \tfrac{1}{2}E_o^2/Z_o$ to the energy flux of the beam. The time-averaged momentum density of the beam (i.e., momentum per unit volume), $p_z = \langle S_z\rangle/c^2$, is thus uniform throughout the free-space region, having equal contributions from the two frequency components of the beam.

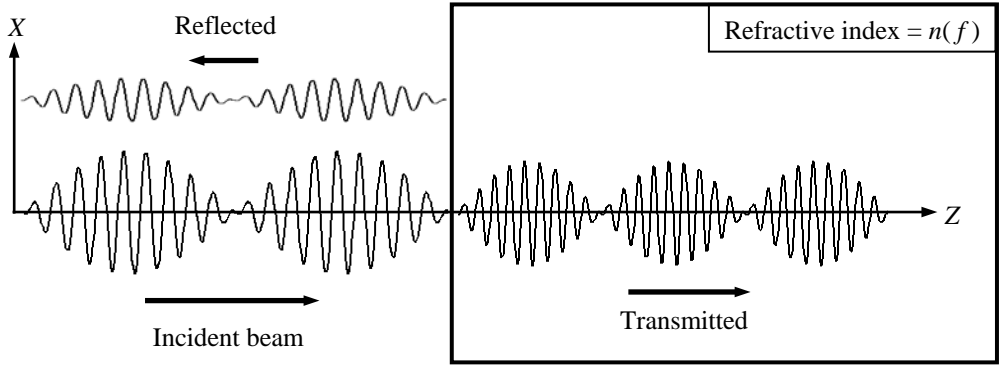

Fig. 1. A light beam consisting of two equal-amplitude plane-waves of slightly differing frequencies, $f_1$ and $f_2$, is normally incident on a semi-infinite dielectric of refractive index $n(f)$. The beam is linearly polarized, having its $E$-field along the $x$-axis and $H$-field along the $y$-axis. While a fraction of the beam is reflected at the surface, the remainder enters the dielectric, penetrating it at the group velocity $V_g = c/(n + n'f)$. Here $f = \tfrac{1}{2}(f_1 + f_2)$ is the center frequency, and $n' = dn/df$; both $n$ and $n'$ are evaluated at the center frequency.

Suppose now that the beam reaches the interface with a dielectric medium of refractive index $n(f)$, as shown in Fig. 1. Each frequency component gets reflected at the interface, with an (amplitude) reflection coefficient $\rho = (1 - n)/(1 + n)$. The time rate of change of the optical momentum $\boldsymbol{q}$ on the vacuum side of the interface is then equal to the rate of arrival of the incident momentum, $cp_z = E_o^2/(cZ_o) = \varepsilon_o E_o^2$, plus the rate of departure of the reflected optical momentum, namely,

$$dq_z/dt = \{1 + \tfrac{1}{2}|(1 - n_1)/(1 + n_1)|^2 + \tfrac{1}{2}|(1 - n_2)/(1 + n_2)|^2\}\varepsilon_o E_o^2. \tag{4a}$$

In the limit $f_1 \to f_2$, the refractive indices $n_1$, $n_2$ in Eq. (4a) are nearly identical, and the above expression simplifies to

$$dq_z/dt = 2[(n^2 + 1)/(n + 1)^2]\varepsilon_o E_o^2, \tag{4b}$$

where $n = \tfrac{1}{2}(n_1 + n_2)$. Momentum conservation requires the above $dq_z/dt$ to be balanced by the force exerted on the dielectric medium plus the time rate of change of any momentum taken by the transmitted beam into the dielectric.

## 3. Superposition of two plane waves in a dielectric medium

Inside the medium, each frequency component arrives with an amplitude transmission coefficient (for the $E$-field) given by $\tau = 2/(1 + n)$. The transmitted $E$- and $H$-fields may thus be written as follows:

$$E_x(z, t) = [2E_o/(n_1 + 1)]\sin\{2\pi f_1[(n_1 z/c) - t]\} - [2E_o/(n_2 + 1)]\sin\{2\pi f_2[(n_2 z/c) - t]\} \tag{5a}$$

$$H_y(z, t) = \{2n_1 E_o/[Z_o(n_1 + 1)]\} \sin\{2\pi f_1 [(n_1 z/c) - t]\}$$
$$- \{2n_2 E_o/[Z_o(n_2 + 1)]\} \sin\{2\pi f_2 [(n_2 z/c) - t]\} \quad (5b)$$

As before, the Poynting vector may be calculated and time-averaged over one beat period $T$. Once again, ignoring the terms in $\Delta f$ and higher order (or assuming $f = N\Delta f$ so that these terms would automatically vanish), we find the average rate of flow of energy (per unit area per unit time) within the dielectric medium to be independent of $z$ and given by

$$<S_z(z,t)> = \tfrac{1}{2}\{[4n_1/(n_1+1)^2]+[4n_2/(n_2+1)^2]\}E_o^2/Z_o. \quad (6)$$

The electromagnetic (Abraham) momentum inside the dielectric whose volumetric density $p_z = <S_z>/c^2$ must propagate with the group velocity $V_g = c(f_2 - f_1)/(n_2 f_2 - n_1 f_1) = c/(n + n'f)$. Here $n = \tfrac{1}{2}(n_1 + n_2)$, $f = \tfrac{1}{2}(f_1 + f_2)$, and $n' = (n_2 - n_1)/(f_2 - f_1)$ is the derivative of $n$ with respect to frequency. In the limit $f_1 \to f_2$, the refractive indices $n_1$ and $n_2$ appearing in Eq. (6) are nearly identical, and the time rate of flow of the electromagnetic momentum through the medium is given by

$$d q_z/dt = p_z V_g = 4n\varepsilon_o E_o^2/[(n+n'f)(n+1)^2]. \quad (7)$$

Clearly, this is *not* equal to the momentum flux across the vacuum-dielectric interface given by Eq. (4b). The reason for the discrepancy is that some of the incoming momentum has been converted into mechanical force exerted on the dielectric medium. To compute this force we use the Lorentz law, taking into account the fact that the polarization density within the medium is $P(z, t) = \varepsilon_o(\varepsilon - 1)E(z, t)$, where $\varepsilon = n^2$ is a function of the frequency $f$. The bound current density is thus given by

$$J_x(z, t) = \partial P_x(z, t)/\partial t = -4\pi f_1 (n_1 - 1)\varepsilon_o E_o \cos\{2\pi f_1 [(n_1 z/c) - t]\}$$
$$+ 4\pi f_2 (n_2 - 1)\varepsilon_o E_o \cos\{2\pi f_2 [(n_2 z/c) - t]\}. \quad (8)$$

The Lorentz force density $F_z(z, t) = J_x \times \mu_o H_y$ is obtained by multiplying Eqs. (8) and (5b). At time $t$ the leading edge of the beat "waveform" (see Fig. 1) has penetrated a distance $z_o = V_g t$ into the dielectric. The integrated force density from $z = 0$ to $z_o$ thus yields the force per unit cross-sectional area, $F_z(t)$, exerted at time $t$ on the medium. We find

$$F_z(t)/(2\varepsilon_o E_o^2) = \{[n_1 n_2 - (n_1 f_2 - n_2 f_1)/(n_2 f_2 - n_1 f_1)]/[(n_1+1)(n_2+1)]\}\{1 - \cos[2\pi(f_2 - f_1)t]\}$$
$$- \{(n_2 - n_1)(f_2 - f_1)/[(n_1 + 1)(n_2 + 1)(n_1 f_1 + n_2 f_2)]\}\cos[4\pi(n_2 - n_1)f_1 f_2 t/(n_2 f_2 - n_1 f_1)]$$
$$+ \{[n_1 n_2 - (n_1 f_2 + n_2 f_1)/(n_1 f_1 + n_2 f_2)]/[(n_1 + 1)(n_2 + 1)]\}\cos[2\pi(f_1 + f_2)t]$$
$$- \tfrac{1}{2}[(n_1 - 1)/(n_1 + 1)]\cos(4\pi f_1 t) - \tfrac{1}{2}[(n_2 - 1)/(n_2 + 1)]\cos(4\pi f_2 t). \quad (9)$$

Averaging $F_z(t)$ over the duration of a single beat waveform, $T = 1/(f_2 - f_1)$, we find that the second, third, fourth, and fifth terms in the above expression contribute very little to the average force, and that the only significant contribution arises from the first term, namely,

$$<F_z> = (1/T)\int_0^T F_z(t)\,dt = 2\varepsilon_o E_o^2 [n_1 n_2 - (n_1 f_2 - n_2 f_1)/(n_2 f_2 - n_1 f_1)]/[(n_1+1)(n_2+1)]. \quad (10)$$

In the limit $f_2 \to f_1$ the above expression simplifies to yield the net (average) force per unit cross-sectional area exerted on the dielectric medium as follows:

$$<F_z> = 2\{n^2 - [(n - n'f)/(n + n'f)]\}\varepsilon_o E_o^2/(n+1)^2. \quad (11)$$

In averaging $F_z(t)$ of Eq. (9) over the time interval $T$, we neglected the contributions of all but the first term. This is easily justified for the third, fourth, and fifth terms, which exhibit rapid oscillations. However, the second term is harder to neglect, especially when $n' \approx 0$ (i.e., for a nearly dispersionless medium), because the cosine term under these circumstances is not rapidly oscillating. However, the magnitude of the second term is proportional to $\Delta n \Delta f$, which reduces the term's significance when the cosine is weakly oscillating. All in all, the time average of the second term in Eq. (9) turns out to be negligible even when the medium is dispersionless (or nearly so).

Adding $<F_z>$ of Eq. (11) to the rate of flow of the electromagnetic (i.e., Abraham) momentum inside the dielectric given by Eq. (7) yields:

$$p_z V_g + <F_z> = 2[(n^2+1)/(n+1)^2]\varepsilon_o E_o^2. \tag{12}$$

This is identical to the total momentum per unit area per unit time imparted to the dielectric as given by Eq. (4b). We have thus confirmed the conservation of momentum in the system of Fig. 1 by deriving the expression for the radiation pressure, Eq. (11), and by requiring propagation at the group velocity $V_g$ for the electromagnetic momentum inside the medium; see Eq. (7).

If mechanical momentum is assumed to travel through the dielectric with the group velocity $V_g$, a mechanical momentum density $p_z^{(mech)} = <F_z>/V_g$ can be defined which, combined with the electromagnetic momentum density $p_z^{(Abraham)} = <S_z>/c^2$, accounts for the total optical momentum residing in the medium. In the limit when $n' \to 0$, for each frequency component of the beam (i.e., $f_1$, $f_2$), $p_z^{(mech)} \to \tfrac{1}{2}(\varepsilon - 1)<S_z>/c^2$, in agreement with the result obtained in [1] for dispersionless media.

## 4. The photon drag effect

An intriguing experimental observation in certain (weakly absorbing) semiconductors, notably Si and Ge, is the photon drag effect [7]. When a photon of energy $hf$ from a monochromatic beam (vacuum wavelength $\lambda_o = c/f$) is absorbed within a semiconductor of refractive index $n$ (i.e., the real part of the complex refractive index $n + i\kappa$, whose imaginary part is the absorption coefficient $\kappa$), the excited charge carrier acquires a momentum equal to $nhf/c$, the so-called Minkowski momentum of the photon. In contrast, the photon's Abraham momentum – seen from the preceding section's discussions to be $hf/[(n+n'f)c]$ – is clearly different from the Minkowski momentum. Combining the photon's electromagnetic and mechanical momenta does not resolve the discrepancy either, as the total photon momentum, $\tfrac{1}{2}[n+(1/n)]hf/c$, differs from the Minkowski momentum as well.

Loudon *et al* [8] have given a comprehensive theory of the photon drag effect, arguing that the "transparent part" of the semiconductor (associated with the real part of the complex refractive index) takes up the difference between the photon's momentum and the Minkowski momentum, when the latter is transferred to the "absorbing part" of the material (i.e., the part associated with the imaginary component of the complex refractive index). We present a similar (though by no means identical) explanation of the photon drag effect by showing that the momentum picked up by a thin absorbing layer embedded in a transparent dielectric is equal to the Minkowski momentum of the incident photon.

With reference to Fig. 2, the reflection and transmission coefficients of an absorbing layer of thickness $d$ and complex index $n + i\kappa$, in the limit of $d \ll \lambda_o$ (where $\lambda_o$ is the vacuum wavelength of the incident beam), can be shown to be

$$\rho = -[1 + i(\kappa/2n)](2\pi\kappa d/\lambda_o), \tag{13a}$$

$$\tau = 1 - (2\pi\kappa d/\lambda_o) + i[2n - (\kappa^2/n)]\pi d/\lambda_o. \tag{13b}$$

The above equations are obtained by assuming the existence of a pair of counter-propagating plane-waves in the absorbing layer, then determining the various unknown amplitudes by matching the boundary conditions for the $E$- and $H$-fields while ignoring second- and higher-order terms in $d/\lambda_o$. (The algebra is straightforward but tedious.) The absorbed optical power $\gamma$ within the layer is thus given by

$$\gamma = \tfrac{1}{2}(1 - |\rho|^2 - |\tau|^2)\, nE_o^2/Z_o = (2\pi n\kappa\, d/\lambda_o)\, E_o^2/Z_o. \tag{14}$$

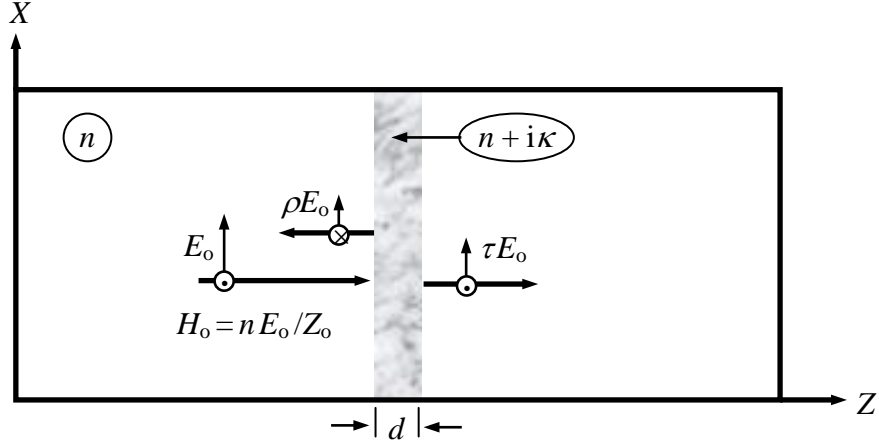

Fig. 2. A thin absorptive layer of thickness $d$ and complex refractive index $n + i\kappa$ is embedded in a transparent, homogeneous dielectric medium of refractive index $n$ (same $n$ as the real part of the complex index of the absorptive layer). A monochromatic plane-wave, having vacuum wavelength $\lambda_o = c/f$, $E$-field amplitude $E_o$, and $H$-field amplitude $H_o = nE_o/Z_o$, is normally incident on the absorbing layer. The layer's (amplitude) reflection and transmission coefficients are $\rho$ and $\tau$, respectively. Each absorbed photon of energy $hf$ transfers the equivalent of its Minkowski momentum $nhf/c$ to the absorbing layer.

The force experienced by the absorbing layer may be derived from the Lorentz law, following the procedure outlined in [1], then integrated over the film thickness $d$ to yield

$$\langle F_z \rangle = (2\pi n^2 \kappa\, d/\lambda_o)\, \varepsilon_o E_o^2. \tag{15}$$

Thus $\langle F_z \rangle = (n/c)\gamma$, namely, the force experienced by the absorbing layer is $(n/c)$ times the captured optical power. Consequently, the momentum transferred to the layer in a given time interval $\Delta t$ must also be $(n/c)$ times the energy absorbed by the layer during the same time interval. For a captured photon of energy $hf$ the momentum transfer is thus equal to $nhf/c$, i.e., the Minkowski momentum of the photon in its dielectric environment. Since this is greater than the total photon momentum prior to being absorbed, the host medium (i.e., the dielectric) must experience a recoil equal to the difference between the incident photon's initial momentum and the Minkowski value picked up by the excited charge carrier.

**Acknowledgments**

The author is grateful to Ewan Wright, Armis Zakharian, Pavel Polynkin, and Walter Hoyer for many helpful discussions. This work has been supported by the AFOSR contract F49620-02-1-0380 with the Joint Technology Office, by the *Office of Naval Research* MURI grant No. N00014-03-1-0793, and by the *National Science Foundation* STC Program under agreement DMR-0120967.